# Giant low frequency velocity fluctuations in a driven vortex lattice

Shyam Mohan, Jaivardhan Sinha, and S. S. Banerjee[a]

Department of Physics, Indian Institute of Technology, Kanpur-208016, India

A. K. Sood[b]

Department of Physics, Indian Institute of Science, Bangalore 560012, India

S. Ramakrishnan and A. K. Grover[c]

Department of Condensed Matter Physics and Materials Science, Tata Institute of Fundamental Research, Mumbai- 400005, India

**The driven state of a well-ordered flux line lattice in a single crystal of 2H-NbSe$_2$ in the time domain has revealed the presence of substantial fluctuations in velocity, with large and distinct time periods (~ seconds). A superposition of a periodic drive in the driven vortex lattice causes distinct changes in these fluctuations. We propose that prior to onset of the peak effect there exist hithertofore unexplored regime of coherent dynamics, with unexpected behaviour in velocity fluctuations.**

---

[a] email: satyajit@iitk.ac.in

[b] email: asood@physics.iisc.ernet.in

[c] email: grover@tifr.res.in

The statistical mechanics of vortices driven through a random pinning environment is a prototype for studying other driven condensed matter systems, like, charge density waves, Wigner crystals[1,2,3,4,5], etc. In recent times an intensely investigated area of study has been the peak effect (PE) regime, where there is an anomalous increase in the critical current, $I_c(H,T)$[6] in the vicinity of the upper critical field ($H_{c2}(T)$). Characteristic behaviour in the voltage (V) response[7,8,9], when magnetic field, temperature and/or transport current are varied; history dependent dynamic response[7,10] and significant increase in the low frequency noise[11,12,13,14,15,16,17], etc. are some of the reported manifestations of PE. Structurally, the vortex state in the PE regime has an admixture of ordered (weakly pinned) and disordered (stronger pinned) phases[18,19,20,21,22,23,24,25]. Dynamic variations in the relative fractions of ordered and disordered phases for the driven vortex matter are considered to be responsible for the observed features in the PE regime[11,12,18]. Absence of reports of such anomalies well before the onset of the PE phenomenon has fortified the notion that the elastic vortex solid phase prior to PE is benign. However, few recent studies have implied other features prior to the PE viz., in the dissipation behaviour for the quasi-static vortex state[26], and also in its response time to a current (drive)[27] pulse. Our present investigations reveal slow temporal velocity fluctuations with characteristic low frequencies, across this elastic regime. These fluctuations evolve on increasing field towards, and across the PE region. Further, we also find nonlinear modulation in the velocity fluctuations in the elastic state by superimposition of a small ac-perturbation. These results call for exploring the nature of vortex dynamics beyond the existing knowledge related to the statistical physics of the driven vortex state.

We chose a superconducting compound 2H-$NbSe_2$ (single crystals with $T_c(0) \sim 7$ K, dimensions $\approx 2.5 \times 1.5 \times 0.1$ mm$^3$), as it typically possess[7] very low $j_c/j_0 \sim 10^{-6}$ ( $j_c$: the critical depinning current density and $j_0$: the de-pairing current density). Isothermal electrical transport measurements were performed using a standard four probe geometry, with the dc magnetic field ($H$) applied parallel to the $c$-axis and the current ($I_{dc}$) along the crystals $ab$ plane. In our sample, the residual resistance ratio ($RRR$), $R$(300 K)/$R$(10 K)[10] is 5. Fig. 1 shows the plots of resistance ($R$) versus $H$ in the single crystal at 4 K (red circles) and 4.5 K (green triangles), respectively. At 4 K, for $H < 2.5$ kOe and dc current $I_{dc}$ = 30 mA, $R < 0.1$ m$\Omega$, which implies an immobile, pinned vortex state. Beyond 2.5 kOe, the FLL gets de-pinned and $R$ increases to m$\Omega$ range, implying the $I_c$ to be 30 mA (at 4 K, 2.5 kOe). The inset (a) in Fig.1 shows that voltage versus current ($V$-$I_{dc}$) at 4 K and 7.6 kOe fits to $V \sim (I_{dc} - I_c)^{\beta}$, where $\beta \sim 2$ and $I_c$ = 18 mA (criterion used: at I = $I_c$, $V$ > 5 $\mu$V). Earlier studies revealed that $\beta > 1$ corresponded to the driven state of a weakly pinned vortex solid prior to the PE[7,19,20]. The differential resistance plot [(d$V$/d$I_{dc}$) vs $I_{dc}$] in the same inset indicates a nonlinear $I$-$V$ response at $I_{dc}$ = 30 mA. Note, however, that the differential resistance (at 7.6 kOe, 4 K) does not display any peak[7,17] feature characteristic of the entry into the PE regime. A threshold field marked $H_{pl}$ in Fig.1 identifies the onset field of the PE region. The inset (b) in Fig.1 shows the phase diagram for the vortex matter driven with $I_{dc}$ = 30 mA in the given crystal for $H \parallel c$. In Fig.1, we designate the region I as the pinned state and the weakly disordered - driven vortex state forms a major part of the region II. The interval, $H_{pl} < H < H_{c2}$ , is divided into two parts (regions III and IV, respectively). The upper (field) limit of the region III identifies the end of the PE, where the pinning in the bulk nearly ceases.

Figures 2 and 3 summarize the results of investigations in time ($t$) domain. The protocol for measurements was as follows: At a given $H$ and $I_{dc}$, the voltage $V_0$ across the sample was recorded by averaging over a large number of measurements (~ 100). To measure the time series, we captured the voltage response ($V(t)$) at $T$ = 4.5 K with a resolution of 35 *ms*. The measurement of $V(t)$ response at a given ($H, T$) is triggered by switching-on $I_{dc}$. We begin recording V(t) after confirming that its time averaged value (<V(t)>) has approached its nominal mean (~$V_0$). The stack of panels (a) to (f) in Fig.2 show $V(t)$ at $T$ = 4.5 K for different driven regimes, viz., (a): the just de-pinned state ($H << H_{pl}$), (b): the weakly disordered - driven vortex state prior to the PE ($H < H_{pl}$), (c): above the onset of the plastic regime ($H > H_{pl}$), and (d) and (e): deep inside the plastic regime ($H \geq$ peak field of the PE) and (f) in the normal state (at $T > T_c$). A striking feature inferred is the fractional change in the V(t) about $V_0$, varying from few percent to few tens of percent of $V_0$ for $H < H_{c2}$, to about 0.02% in the normal state (e.g. at 10 K and 10 kOe). The noise level in the normal state was ~ $10nV/\sqrt{Hz}$. Such conspicuously large amplitude slow fluctuations about $V_0$ sustained prior to the PE (upto 7.6 kOe), begin to degrade on entering the PE regime (i.e., see panels (c) and (d) in Fig.2). To overcome the influence of the inevitably present random electronic noise, the panels (g) to (l) show the voltage-voltage correlation function, $C(t)$ ($= \frac{1}{V_0^2}\langle V(t'+t)V(t')\rangle$), calculated from the $V(t)$ signals in Fig 2. The $C(t)$ plots in panels (g) to (l) in Fig.2 reveal interesting velocity correlations in time, viz., distinct nearly periodic fluctuations at *7.6 kOe* ($H < H_{pl}$ in panel (h)) shift in frequency at $H > H_{pl}$ viz., at 10.8 kOe (cf. panel (i)). The C(t) in the normal state (Fig.2(l)) is featureless.

The power spectrum was deduced via the fast Fourier transform (FFT) of $C(t)$, determined numerically, see panels (m) to (r) in Fig.2. At 2.6 kOe, for the just depinned vortex array at $I_{dc}$ = 30 mA, there are two peak-like features, centered around 0.25 Hz and 2 Hz (Fig.2(m)). For the driven state of the weakly disordered vortex solid prior to the PE at 7.6 kOe, a distinct sharp peak develops close to 0.25 Hz (Fig.2(n)), with an amplitude nearly five times that at 0.25 Hz for 2.6 kOe. Within the plastic regime ($H > H_{pl}$) at 10.8 kOe, the strongest peak is close to 2 Hz (Fig.2(o)). Beyond the peak position of PE (e.g., at 13.6 kOe), where the pinning has almost vanished, the peak at ~ 0.25 Hz is the most dominant feature (cf. Fig. 2(p)). In Fig.2(q), which correspond to 16 kOe, the bands at 0.25 Hz and 0.5 Hz (though weak in amplitude), are clearly identifiable. The continuing presence of a significant fraction of the weakly pinned phase of vortices in a dynamically coexisting two phase mixture across the entire PE regime[18] could rationalize the position of the peak at ~ 0.25 Hz in the noise power spectra. Also, significant to note, the noise power spectrum in the normal state is featureless (cf. Fig.2(r) and Fig.2(n)). In the parameter regime investigated, we determined from the power spectra of temperature fluctuations (data not shown here) recorded at different H that the temperature fluctuations were random and did not correlate with the characteristic peaks in the voltage spectra and thus do not account for their presence and/or evolution.

The observation that the low frequency mode(s) are excited in the driven (by $I_{dc}$) vortex solid prior to the PE, led us to explore the effect of an ac current ($I_{ac}$) superimposed on $I_{dc}$. Fig.3 shows the measured dc voltage response ($V_{dc}$) for a current, $I = I_{dc} + I_{ac}$, where $I_{dc}$ = 22 mA and $I_{ac} = I_0\mathrm{Cos}(2\pi ft)$, with $I_0$ = 2.5 mA. The frequency ($f$) was varied in discrete steps in Fig.3. The mean $V_{dc}$ level of about 8 μV in Fig.3 corresponds to $I_{dc}$ of 22 mA (compare this with inset (a) in Fig.1). The observation of large (~100%) excursions in the measured $V_{dc}$ signal at harmonics of 0.25 Hz indicates a significantly large *nonlinear* response in the traditionally assumed *linear*, weakly disordered - driven vortex solid prior to the PE. The features such as in Fig.3 progressively degrade as the PE sets in and field further enhances into the PE regime.

We propose that the resistance of the sample varies as $R = R_0 \pm \sum_{n=1}^{m}[R_n Cos(n2\pi f_0 t) + R_n^{/} Sin(n2\pi f_0 t)]$, under the influence of current, $I = I_{dc} + I_{ac}$.

Here, $R_0$ is the resistance of the sample in response to the $I_{dc}$ alone, $R_n$ and $R_n^{/}$ are the $f$ dependent coefficients of the in-phase and out-of-phase responses, and $f_0$ is the characteristic frequency of fluctuations. The $f_0$ (= 0.25 *Hz*) corresponds to the peak value in the power spectrum for $H$ = 7.6 kOe and $T$ = 4.5 K in the panel (n) in Fig.2. Taking the time average of the expression, $V = IR$, yields, $V_{dc} = I_{dc}R_0 \pm (I_0R_n)/2$, at $f = nf_0$. From the very large fluctuations (~100%) seen in Fig.3, it is clear that $(I_0R_1)/2 \approx I_{dc}R_0$ or $R_1 \sim 20\ R_0$, is a substantially large component excited at $f = f_0$. Similarly, at $f = 2\ f_0$, $R_2 \sim 15\ R_0$. Notice from Fig.3, that the nonlinear response can be easily seen upto $f = 5\ f_0$ (see the positions of black arrows in Fig.3). The envelope of the amplitude of fluctuations in Fig.3 appears to decrease upto 5 $f_0$; thereafter, the envelope regenerates itself into second and third cycles of oscillations, but, with progressively, reduced intensities. A maximum in the envelope of the amplitude is also present at about 2 Hz. Thus, a small perturbation with $I_{ac}$ ~ 0.1 $I_{dc}$ triggers large fluctuations along with a higher-harmonic generation indicating a highly nonlinear nature of the dynamics.

While above features are absent, we find additional features in V(t) appearing at the onset of the PE. By capturing V(t) with a higher time resolution of 1.25 ms, we find the onset of PE shows an interesting intermittency behavior in the fluctuation. The inset (a) in Fig.3 shows intermittent large voltage (≡ velocity) bursts, which are almost twice as large as the mean level of ~ 150 μV at $T$ = 6 $K$ (i.e., closer to $T_c(0)$) and near the onset field of the PE at that temperature (i.e., at 2.2 kOe). Such bursts are followed by time intervals, when the fluctuations are nearly periodic. Similar fluctuations are found in the simulations reported by Olive and Soret (OS)[17] for the driven plastically deformed vortex state. OS[17] proposed that the above intermittent fluctuations signal the onset of a chaotic motion, wherein intermittently, the vortex motion within mobile vortex channels periodically synchronizes with the fluctuating vortices trapped in the pinned islands leading to periodic fluctuations. We assert that we have uncovered in the time series data an evidence for substantial nonlinear fluctuations even in the preceding region II preceeding PE along with the onset of chaotic motion in the driven plastic phase in region III.

In the driven vortex matter velocity fluctuations exist in the range of the washboard frequency[28] ~ 0.1-1 MHz, which arises when periodically spaced vortices are driven over pins. The appearance of a large enhancement in the nonlinear *I-V* characteristics across the PE region[7] with low frequency noise in the range of few Hz[7,11,12,13,15,16] (<<

washboard frequency) has been analyzed invoking the edge contamination model[11,12]. In our case, vortices need about 0.1 s to traverse the typical width of our sample of ~ 0.1 cm, with a vortex velocity , $u = \langle V(t)\rangle/(d.B) \sim 10^{-2}$ m/s( = 1 cm/s), where $V \sim 10$ μV observed at 30 mA, $B = \mu_0 H = 1$ T, and $d$ is the distance between the electrical contacts = $10^{-3}$ m. Therefore, the injection rate of disordered vortices into the moving vortex medium from irregularities at the sample edges is ~ 10 Hz, which via the edge contamination model should result in velocity fluctuations in a frequency range close to 10 Hz. In the PE region the observed peak in the fluctuation spectrum is centered around 2 Hz (cf. Fig.2(o)) which could be termed as consistent with earlier observations of peak in noise power in similar frequency range in the PE regime of $NbSe_2$ [Ref.12] and $YBa_2CuO_{7-\delta}$ [Ref.14]. Paltiel et al.[11,12] had rationalized the peak in the noise power density at 3 Hz in the PE regime via the edge contamination framework. However, in the weakly disordered – driven vortex state prior to PE (the so called elastic vortex state), we also observe a much lower frequency of 0.25 Hz (cf. Fig.2 (g)), which as per the edge contamination model would imply an effective sample width of 4 cm (with $u = 1$ cm/s) >> actual sample width (~ 0.1 cm). We also show in the inset (b) of Fig.3 an evolution in these characteristic fluctuations frequencies with velocity (u) of the vortices. Significant variation in <V(t)> and hence $u$ is obtained by performing V(t) measurements at different T (i.e., higher T) and H. From a conventional noise mechanism, e.g., edge contamination model, one would expect the higher frequency should increase with u (equivalent to the disorder injection rate) without showing any tendency towards saturation. However, this is not the case here as seen in the inset (b) of Fig.3. The lower frequency appears to be nominally velocity independent. We may clarify that in certain velocity regimes only one of the two frequencies survives. We believe that the detailed richness of the fluctuations in our data does not find a rationalization within the present formulation of the edge contamination model[11,12].

At the end, we propose that the nonlinear response deep within the driven elastic medium is related to a possible transformations in the vortex matter observed deep within the elastic phase[26,27]. Complex nonlinear systems under certain conditions can produce slow spontaneous organization in its dynamics. Under the influence of a sufficient driving force, the system can exhibit coherent dynamics, with well-defined one or more frequencies[29]. The evolution of fluctuations, such as those illustrated in Fig.2, can be viewed as the complex behavior of a nonlinear driven vortex state with multiple attractors (stable cycles). Underlying phase transformations in the driven vortex state induce the system to fluctuate between different stable cycles, leading to a typical spectrum of fluctuations discussed in Fig.2. The above nature may lead to extreme sensitivity of the driven vortex system to the low amplitude perturbations, as is shown in Fig.3.

In conclusion, we have investigated the nature of vortex velocity fluctuations in the time domain and have uncovered signatures of complex nonlinear dynamics. We find new regimes of coherent driven dynamics in the elastic phase with distinct frequencies of fluctuation. These regimes are a precursor to chaotic fluctuations we observe deep in the plastic regime. We hope that the results presented will motivate the need to look beyond existing models to unravel the nature of organization of vortex trajectories and the flow within the elastically driven vortex medium prior to PE.

We thank Mr. M. J. Higgins for the single crystal samples of $2H\text{-}NbSe_2$ grown at NEC Research Institute, Princeton, U.S.A.. We are grateful for the experimental support from Dr. Ajay Thakur and Mr. Ulhas Vaidya. SSB thanks Prof. Eli Zeldov for illuminating discussions pertaining to the edge contamination model. SSB would also like to acknowledge the financial support from DST India, CSIR India and IIT-Kanpur. AKS thanks CSIR for the Bhatnagar Fellowship.

References:

[1] G. Blatter, M. V. Feigel'man, V. B. Geshkenbein, A. I. Larkin, and V. M. Vinokur, Rev. Mod. Phys. **66**, 1125 (1994).

[2] T. Nattermann and S. Scheidl, Adv. Phys. **49**, 607 (2000).

[3] T. Giamarchi and P. Le Doussal, Phys. Rev. Lett. **76**, 3408 (1996).

[4] S. Scheidl and V. M. Vinokur, Phys. Rev. E **57**, 2574 (1998).

[5] T. Giamarchi and S. Bhattacharya, in High Magnetic fields: Applications to Condensed Matter Physics and Spectroscopy, edited by C. Berthier, L. P. Levy and G. Martinez (Springer Verlag, 2002), p.314.

[6] P. W. Anderson, Basic Notions in Condensed Matter Physics, Addison – Wesley, New York, U. S. A., pp.162-163, 1983.

[7] S. Bhattacharya and M. J. Higgins Phys. Rev. Lett. **70**, 2617 (1993); M. J. Higgins and S. Bhattacharya, Physica C **257**, 232 (1996).

[8] M. C. Faleski, M. C. Marchetti, and A. A. Middleton, Phys. Rev. B **54**,12427 (1996).

[9] A. E. Koshelev and V. M. Vinokur, Phys. Rev. Lett. **73**, 3580 (1994).

[10] S. S. Banerjee *et al.*, Phys. Rev. B **59**, 6043 (1999); S. S. Banerjee *et al*., Physica C **308,** 25 (1998).

[11] Y. Paltiel *et al*., Nature **403**, 398 (2000).

[12] Y. Paltiel *et al*., Europhys. Lett. **58**, 112 (2002).

[13] S. N. Gordeev *et al.,* Nature **385**, 324 (1997).

[14] G. D. Anna, P. L. Gammel, H. Safar, G. B. Alers, and D. J. Bishop, Phys. Rev. Lett. **75**, 3521 (1995).

[15] R. D. Merithew, M. W. Rabin, M. B. Weissman, M. J. Higgins, and S. Bhattacharya, Phys. Rev. Lett. **77**, 3197 (1996); M. W. Rabin, R. D. Merithew, M. B. Weissman, M. J. Higgins, and S. Bhattacharya, Phys. Rev. B **57**,R720(1998)..

[16] A. C. Marley, M. J. Higgins, and S. Bhattacharya, Phys. Rev. Lett. **74**, 3029 (1995).

[17] E. Olive and J. C. Soret, Phys. Rev. Lett. **96**, 027002 (2006); *ibid* Phys. Rev. B **77**, 144514 (2008).

[18] M. Marchevsky, M. J. Higgins, and S. Bhattacharya, Nature **409**, 591 (2001).

[19] U. Yaron *et al*., Nature **376**, 753 (1995).

[20] A. Duarte *et al*., Phys. Rev. B **53**, 11336 (1996).

[21] A. M. Troyanovski, J. Aarts, and P. H. Kes, Nature **399**, 665 (1999).

[22] T. Matsuda, K. Harada, H. Kasai, O. Kamimura, and A. Tonomura, Science **271**, 1393 (1996).

[23] F. Nori, Science **271**, 1373 (1996).

[24] Y. Abulafia *et al*., Phys. Rev. Lett. **77**, 1596 (1996).

[25] C. Reichhardt, C. J. Olson, and F. Nori, Phys. Rev. B **58**, 6534 (1998).

[26] S. Mohan, J. Sinha, S. S. Banerjee, and Y. Myasoedov, Phys. Rev. Lett. **98**, 027003 (2007).

[27] G. Li, E. Y. Andrei, Z. L. Xiao, P. Shuk, and M. Greenblatt, Phys. Rev. Lett. **96**, 017009 (2006).

[28] A. T. Fiory, Phys. Rev. Lett. **27**, 501 (1971); R. M Fleming & C. C. Grimes, Phys. Rev. Lett. **42**, 1423 (1979); N. Kokubo *et al*., Phys. Rev. Lett. **95**, 177005 (2005).

[29] Rajesh Ganapathy and A. K. Sood, Phys. Rev. Lett. **96**, 108301 (2006); R Ganapathy, S Mazumdar, A K Sood, Phys Rev E **78**, 21504(2008).

Figure captions:

**Figure 1.** $R(H)$ curves at 4 K(red circles) and 4.5 K(green triangles) measured with $I_{dc}$ =30 mA and for $H$ || c in 2H-$NbSe_2$. Inset (a) shows $V$-$I_{dc}$ characteristics (open circles) and $dV/dI_{dc}$ versus $I_{dc}$ (filled squares) at $T$ = 4 K and $H$ = 7.6 kOe and the red line represents the fit, $V \sim (I_{dc}\text{ (in mA)} - 18)^2$. Inset (b) displays vortex phase diagram in the given crystal at $I_{dc}$ =30 mA.

**Figure 2.** Panels (a) to (f) represent the fluctuations in voltage $V(t)$ measured at different fields at 4.5 K after $I_{dc}$ of 30 mA was switched on. Panels (g) to (l) in Fig.2 are the voltage – voltage correlation function ($C(t)$) calculated (see text for details) from $V(t)/V_0$ corresponding to the panels (a) to (f), where $V_0$ represents the voltage measured by averaging over a hundred readings at fixed H and T and $I_{dc}$. Panels (m) to (r) show the amplitude of the FFT spectrum calculated from the *C(t)* in panels (g) to (l) (see text for details).

**Figure 3.** $V_{dc}$ versus frequency (*f*) for $I_{dc}$ (= 22 mA) + $I_0$ (= 2.5 mA) $Cos(2\pi ft)$. The $I_{ac}$ + $I_{dc}$ was chosen so that the mean voltage level was comparable to that without $I_{ac}$ and $I_{dc}$ = 30 mA at 7.6 kOe. Inset (a) shows V(t) measured with time resolution of 1.25 ms, at onset of PE at 6K and 2.2 kOe. Inset (b) shows the evolution of the characteristic frequencies with velocity of vortices (u) (see text for details)).

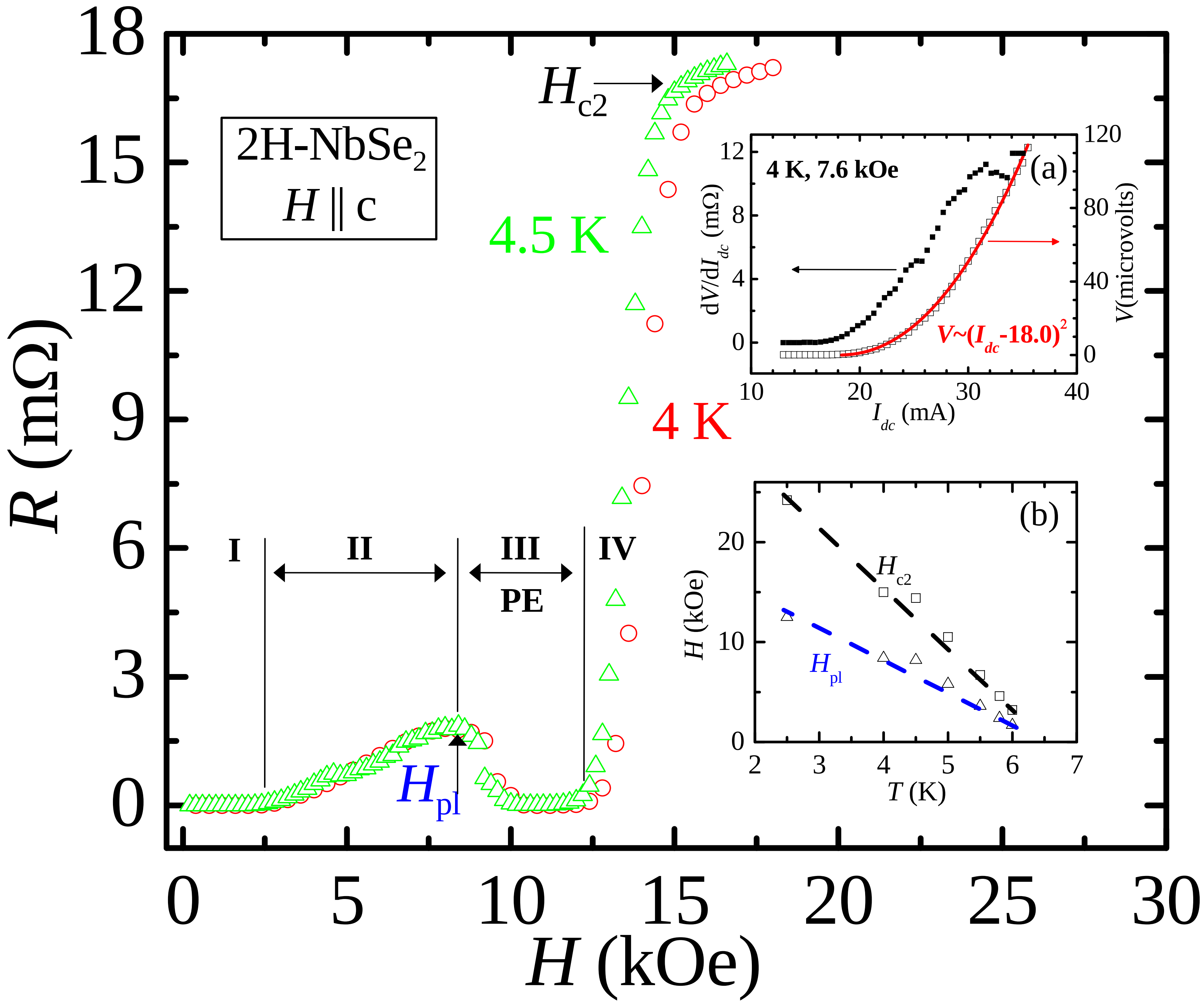

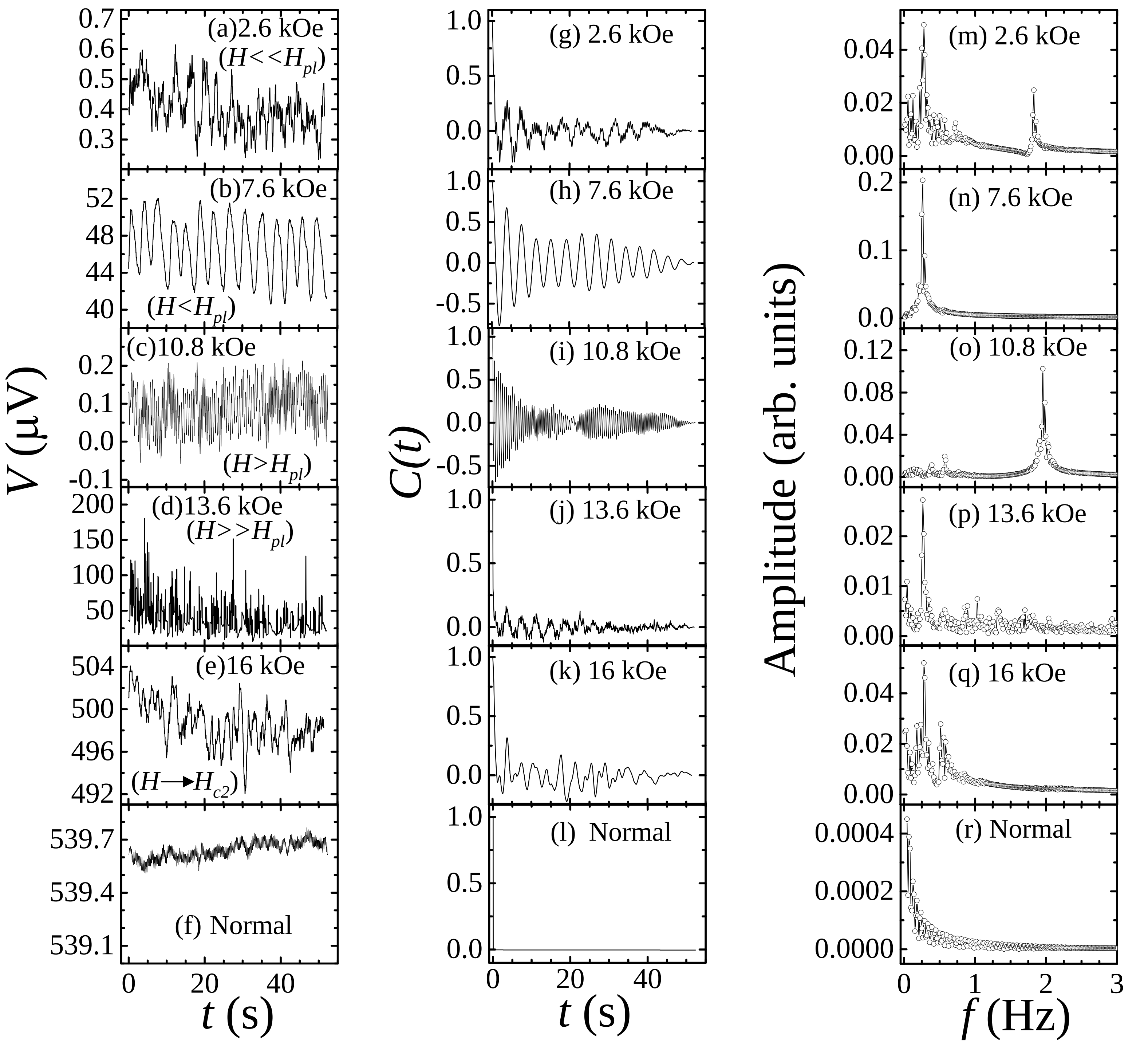

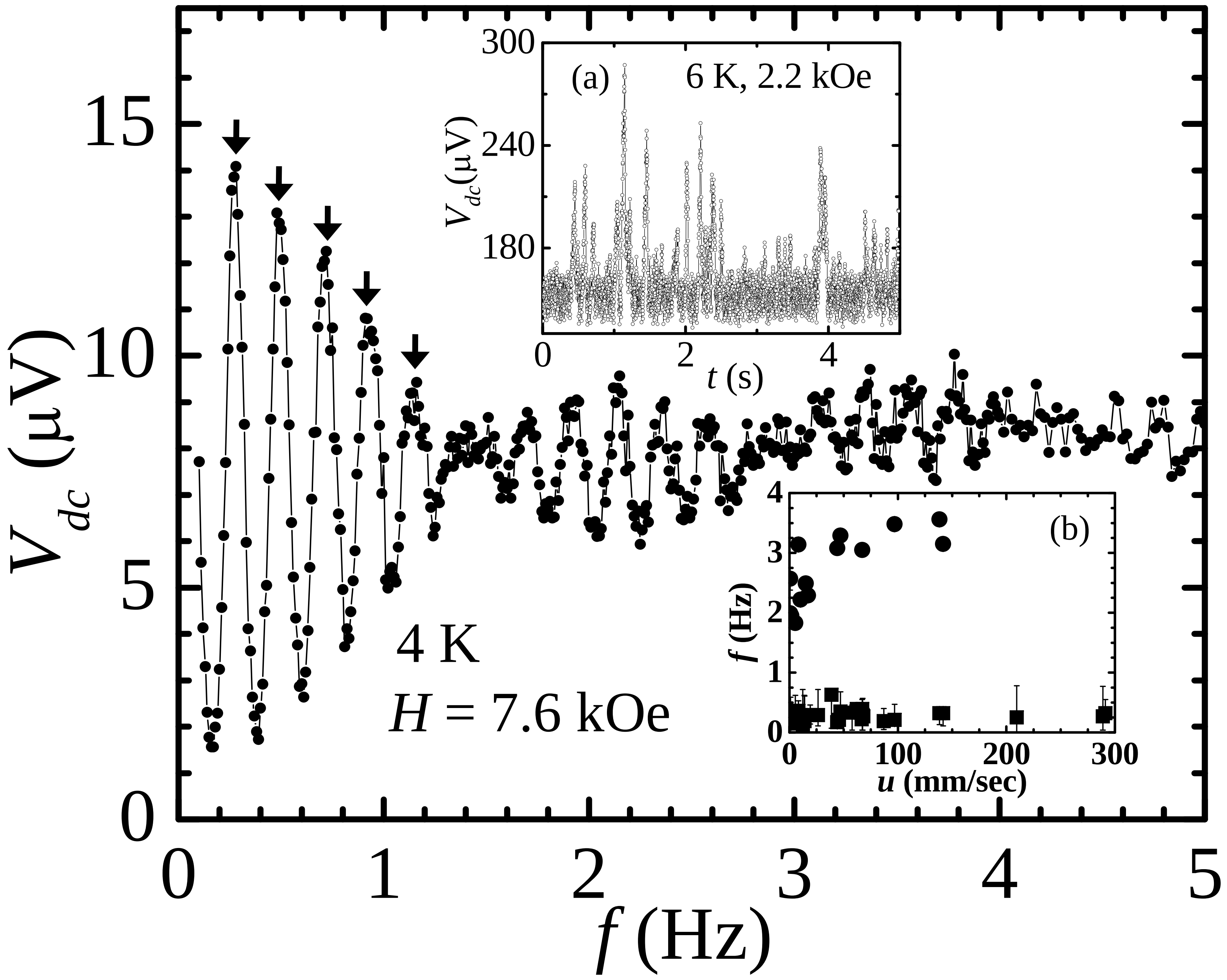